\documentclass[manuscript]{aastex}

\usepackage{graphicx}
\usepackage{fixltx2e} % for superscript, subscript text: \textsuperscript{} & \textsubscript{}
\usepackage{dcolumn}
\usepackage[version=3]{mhchem} % for chemical formula: e.g. \ce{H2O}, \ce{C^{18}O}
\usepackage{natbib}
\usepackage{color}
\usepackage{latexsym}
\usepackage{epstopdf}
\usepackage{rotating}

\newcommand{\kms}{km\,s$^{-1}$}
\newcommand{\Kkms}{K\,km\,s$^{-1}$}

\newcommand{\solarM}{M$_\odot$}

\newcommand{\ci}{[C {\sc i}]} % [CI] line emission
\newcommand{\atci}{C {\sc i}} % CI - atomic carbon
\newcommand{\hii}{H {\sc ii}}

\slugcomment{}

\shorttitle{CI, CO and H2 column density}
\shortauthors{Lo et al.}

\begin{document}


\title{Tracing H$_2$ column density with atomic carbon (C I) and CO isotopologs}


\author{N. Lo} %\altaffilmark{1}}
\affil{Departamento de Astronom\'ia, Universidad de Chile, Camino El Observatorio 1515, Las Condes, Santiago, Casilla 36-D, Chile}

\author{M. R. Cunningham and P. A. Jones}
\affil{School of Physics, University of New South Wales, Sydney, NSW 2052, Australia}

\author{L. Bronfman}
\affil{Departamento de Astronom\'ia, Universidad de Chile, Camino El Observatorio 1515, Las Condes, Santiago, Casilla 36-D, Chile}

\author{P. C. Cortes}
\affil{Joint ALMA Observatory, Santiago, Chile}

\author{R. Simon}
\affil{Physikalisches Institut, Universit{\"a}t zu K{\"o}ln, Z{\"u}lpicher Stra{\ss}e 77, 50937 K{\"o}ln, Germany}

\author{V. Lowe}
\affil{School of Physics, University of New South Wales, Sydney, NSW 2052, Australia}
\affil{Australia Telescope National Facility, CSIRO Astronomy and Space Science, PO Box 76, Epping, NSW 1710, Australia}

\author{L. Fissel and G. Novak}
\affil{Northwestern University, Center for Interdisciplinary Exploration and Research in Astrophysics (CIERA) and Department of Physics \& Astronomy, 2145 Sheridan Road, Evanston, IL 60208, USA }


%\altaffiltext{1}{FONDECYT Postdoctoral Fellow}


\begin{abstract} 
We present first results of neutral carbon ([C I] $^3P_1-^3P_0$ at 492 GHz) and carbon monoxide (\ce{^{13}CO}, J = 1 - 0) mapping in the Vela Molecular Ridge cloud C (VMR-C) and G333 giant molecular cloud complexes with the NANTEN2 and Mopra telescopes. For the four regions mapped in this work, we find that [C I] has very similar spectral emission profiles to \ce{^{13}CO}, with comparable line widths. We find that [C I] has opacity of 0.1 - 1.3 across the mapped region while the [C I]/\ce{^{13}CO} peak brightness temperature ratio is between 0.2 to 0.8. The [C I] column density is an order of magnitude lower than that of \ce{^{13}CO}. The \ce{H2} column density derived from [C I] is comparable to values obtained from \ce{^{12}CO}. Our maps show C I is preferentially detected in gas with low temperatures (below 20 K), which possibly explains the comparable \ce{H2} column density calculated from both tracers (both C I and \ce{^{12}CO} under estimate column density), as a significant amount of the C I in the warmer gas is likely in the higher energy state transition ([C I] $^3P_2-^3P_1$ at 810 GHz), and thus it is likely that observations of both the above [C I] transitions are needed in order to recover the total H$_2$ column density.
\end{abstract}





\section{Introduction}
%\noindent \textit{abstract: no more than 250 words\\
%main text: no more than 3500 words\\
%figures and tables: no more than 5 in total\\
%maximum of 9 sub-figures in one figure\\
%figures can be in eps or pdf: labelled as f1.eps, f2a.eps, f2b.eps, etc\\
%ref: no more than 50\\
%}
%
Carbon monoxide (\ce{CO}) is often used as a tracer of \ce{H2} density in molecular clouds, as it is abundant and easy to observe. However, it is known to be unreliable as it is optically thick in star forming regions \citep[e.g.][]{Shetty2011}. With a new generation of telescopes capable of mapping at sub-millimeter wavelengths, there is a revival of interest in utilizing neutral atomic carbon as a tracer of molecular cloud column density, both observationally and in numerical simulations \citep[see e.g.][]{Shimajiri2013,Offner2014,Glover2014}. One of the perceived advantages of atomic carbon (\atci) over \ce{CO} is that it continues to exist in regions of low dust extinction and strong radiation (commonly found in star forming regions), where \ce{CO} suffers from photodissociation and turns into neutral carbon and oxygen. Here we investigate the effect that density and temperature have on \ce{H2} column density calculated from \ci\ and how it compares to that calculated from the \ce{CO} isotopologues.

Of the four sources mapped here (see Table \ref{tab:col}), three are in the G333 GMC, located at a distance of 3.6 kpc, while NE-RCW36 is north-east of the RCW36 \hii\ region in the Vela Molecular Ridge cloud C (VMR-C) at a distance of 700 pc. Both the G333 and VMR-C clouds have been extensively mapped at various wavelengths, including IR (Spitzer GLIMPSE and MIPSGAL, BLAST, Herschel), millimeter dust, and molecular lines \citep[e.g.][]{Yamaguchi1999,Mookerjea2004,Wong2008,Lo2009,Netterfield2009}. From 1.2 mm dust continuum observations, the sources (clumps) in G333 have masses of $\sim10^3$ \solarM\ \citep{Mookerjea2004}, while the clumps in NE-RCW36 have an order of magnitude lower mass of $\sim70$ \solarM\ \citep{Netterfield2009}. Over the last two years, we have started a program to map these two clouds in atomic carbon emission, with the NANTEN2 Telescope in Chile. We aim to investigate the distribution of \atci\ in comparison to not only the \ce{CO} isotopologues, but also various other chemical species such as \ce{N_2H+}, \ce{HNC}, \ce{CS}, and other dense gas tracers between 86 and 99 GHz. 

We are currently completing the mapping of \atci\ in G333 and the VMR-C, and here present our first result from the four mapped regions, focusing on the comparison between \atci\ and \ce{CO} isotopologues. The selected regions are all associated with embedded stellar clusters and \hii\ regions (see Table \ref{tab:col}).

\section{Observations and data reduction}\label{sec:observations}

%% In a manner similar to \objectname authors can provide links to dataset
%% hosted at participating data centers via the \dataset{} command.  The
%% second curly bracket argument is printed in the text while the first
%% parentheses argument serves as the valid data set identifier.  Large
%% lists of data set are best provided in a table (see Table 3 for an example).
%% Valid data set identifiers should be obtained from the data center that
%% is currently hosting the data.
%%
%% Note that AASTeX interprets everything between the curly braces in the 
%% macro as regular text, so any special characters, e.g. "#" or "_," must be 
%% preceded by a backslash. Otherwise, you will get a LaTeX error when you 
%% compile your manuscript.  Special characters do not 
%% need to be escaped in the optional, square-bracket argument.

The atomic carbon (\ci\ ($^3P_1-^3P_0$) at 492.16 GHz, hereafter simply referred to as \ci) observations presented in this paper were carried out in 2013 with the NANTEN2 Telescope\footnote{http://www.astro.uni-koeln.de/nanten2/} at Pampa la Bola, Chile. The maps were collected with the $2\times8$ pixel array KOSMA SMART receiver and a eXtended bandwidth Fast Fourier Transfer Spectrometer (XFFTS) as backend, 2.5 GHz bandwidth and 32768 channels. At 465 GHz it has a full width to half-maximum (FWHM) beam size of $\sim37$ arcseconds. Data were reduced with {\sc CLASS} ({\sc GILDAS} package), and have velocity resolution of 0.2 \kms. 

The \ce{CO}, \ce{^{13}CO} and \ce{C^{18}O} maps were collected with the Mopra Telescope throughout 2005 to 2007 for G333, 2012 and 2014 for VMR-C \citep[for details see][]{Bains2006,Wong2008,Lo2009}. At 3 mm wavelengths the Mopra telescope has a FWHM beam size of $\sim$36 arcseconds and velocity resolution of $\sim0.1$ \kms\ channel$^{-1}$ at 100 GHz \citep{Ladd2005}. Data were reduced using the {\sc Livedata} and {\sc Gridzilla} packages from the CSIRO/CASS. The Mopra data  are gridded to the same spacing as the NANTEN2 \ci\ data, which equates to 0.13 pc for RCW36 and 0.64 pc for the three G333 sources.

%% In this section, we use  the \subsection command to set off
%% a subsection.  \footnote is used to insert a footnote to the text.


%% This section also includes several of the displayed math environments
%% mentioned in the Author Guide.

\section{Results}
\subsection{Spatial and velocity distributions}\label{sec:maps_vel}
\begin{figure*}
  \epsscale{0.9}
  \plotone{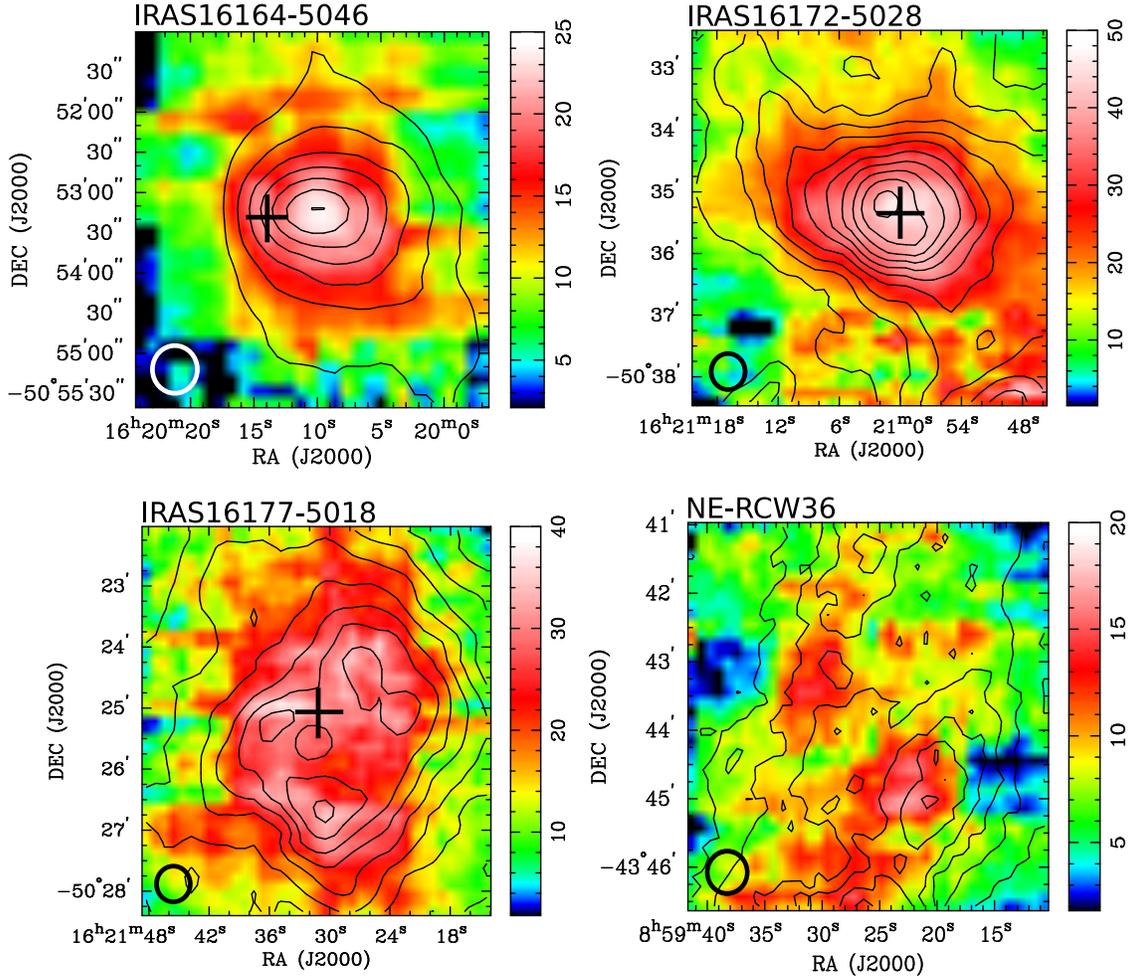}
  \caption{Integrated emission maps of \ce{^{13}CO} (contours) overlaid on \ci\ of the four sources. The \ce{^{13}CO} contour levels start at 30 per cent of the peak integrated emission, with increments of $10\sigma$ for NE-RCW36, and $20\sigma$ for the other three sources. \ce{^{13}CO} peak integrated emission are 131, 189, 164 and 28 \Kkms\ for IRAS16164--5046, IRAS16172--5028, IRAS16177--5018 and NE-RCW36 respectively. $1\sigma$ level of \ce{^{13}CO} integrated emission maps are 0.7 \Kkms\ for IRAS16164--5046, IRAS16172--5028 and IRAS16177--5018, 0.3 \Kkms\ for NE-RCW36, $3\sigma$ level are outside the presented \ce{^{13}CO} maps. Color scales are clipped at $3\sigma$ level, where $1\sigma$ is 0.7 \Kkms\ for IRAS16164--5046 and IRAS16172-5028, 0.6 \Kkms\ for IRAS16177-5018 and NE-RCW36. Beam sizes are marked by circles. The crosses mark the location of the {\it IRAS} sources, except RCW36 which is outside the map.\label{fig:mom0_maps}}
\end{figure*}

\begin{figure}
  \epsscale{0.9}
  \plotone{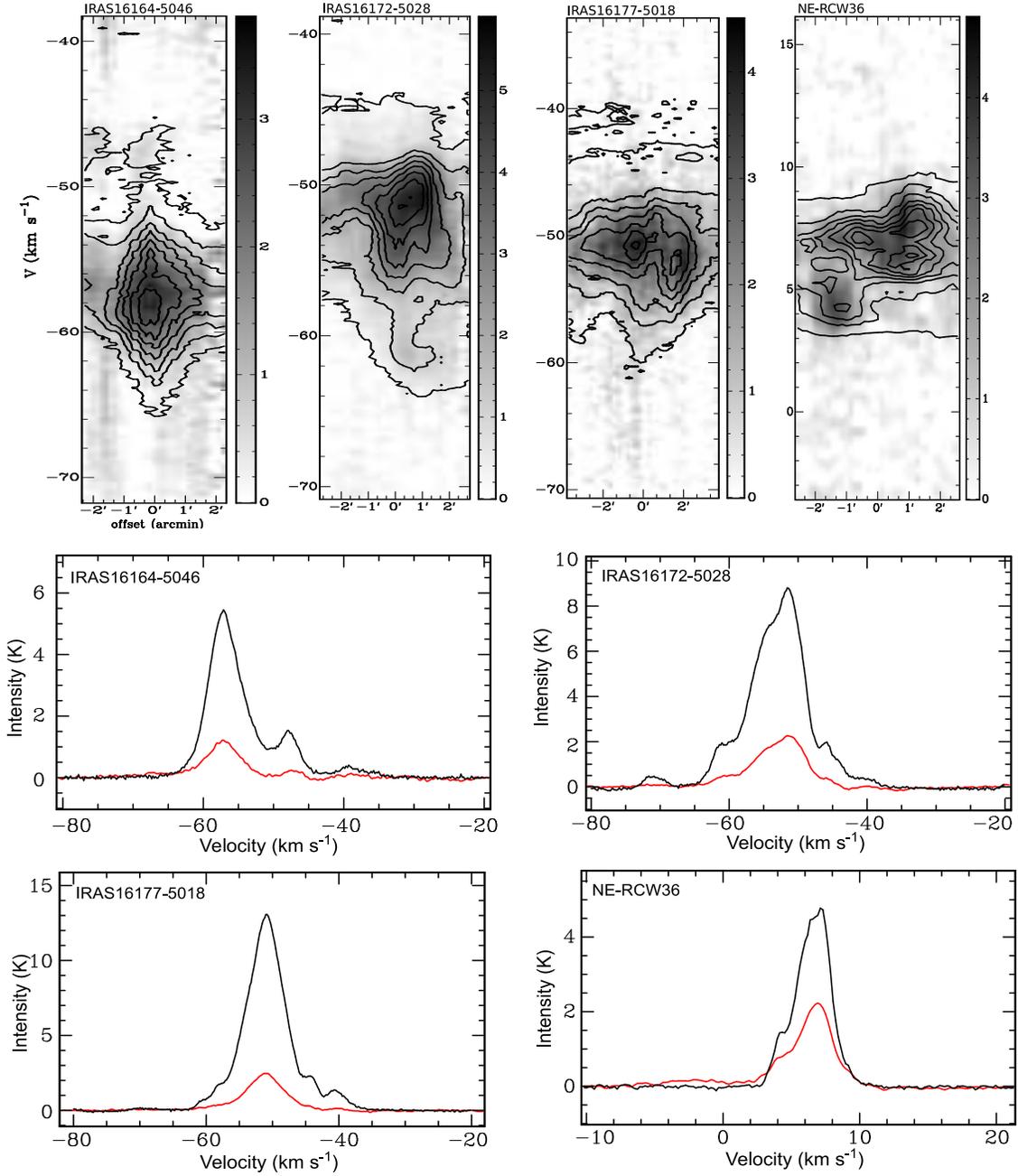}
  \caption{({\it Top row}) \ci\ and \ce{^{13}CO} (contours) position-velocity (PV). The contour levels start at $5\sigma$ noise level with increment of $10\sigma$, where $1\sigma$ is 0.2 K for IRAS16164--5046, 0.3 K for IRAS16172--5028 and IRAS16177--5018, and 0.1 K for NE-RCW36. PV cuts across each of the sources from NE to SW. ({\it Line profiles}) Averaged \ci\ (red) and \ce{^{13}CO} (black) spectra over the area shown in Figure \ref{fig:mom0_maps}.\label{fig:pv_maps_spec}}
%  \label{fig:ratio_maps_spec}}
\end{figure}

Integrated emission maps of \ce{^{13}CO} (contours) overlaid on \ci\ (color) of the four sources are shown in Figure \ref{fig:mom0_maps}. Integrated velocity ranges are $-70$ to $-52$ \kms\ for IRAS16164--5046, $-65$ to $-33$ \kms\ for IRAS16172--5028 and IRAS16177--5018, and 2 to 12 \kms\ for NE-RCW36. There are two distinct velocity components in IRAS16164--5046 (Figure \ref{fig:pv_maps_spec} spectra), $-57$ \kms\ is the main component while the $-48$ \kms\ component is due to a source south-east of IRAS16164--5046 and is outside the presented map. Hence the intensity map and calculations follow are based on emission integrated over the $-57$ \kms\ velocity component for this source. 

In general the bulk of \ci\ emission for the three G333 sources has a similar distribution to the \ce{^{13}CO} emission, with its peak integrated emission also coinciding well with \ce{^{13}CO}, such as the ring-structure seen in IRAS16177--5018 (Figure \ref{fig:mom0_maps} bottom left). In contrast for NE-RCW36 (Figure \ref{fig:mom0_maps} bottom right) there are two \ci\ peaks, the north-east peak (${\rm RA=9^h59^m32^s,Dec=-43^d43^m50^s}$) does not have any corresponding \ce{^{13}CO} emission peak, and the south-west one (${\rm RA=9^h59^m23^s,Dec=-43^d45^m00^s}$) is offset from the \ce{^{13}CO} clump (contours).

% Integrated emission maps

To show the velocity structure of these regions, position-velocity (PV) diagrams of \ci\ (grey scale) and \ce{^{13}CO} (contours) are plotted in Figure \ref{fig:pv_maps_spec} (top row), with line profiles of \ci\ and \ce{^{13}CO} averaged over the region in the integrated emission maps (Figure \ref{fig:mom0_maps} bottom four). \ci\ and \ce{^{13}CO} have similar velocity structure, in the form of centroid velocity, line widths, line wings/shoulders due to outflows and infall, as show in the Gaussian fitted parameters of the spectra (Table \ref{tab:col}). As mentioned previously, the $-48$ \kms\ component of IRAS16164--5046 is due to a different source outside the presented map; here the PV diagram shows that this emission is detached from the main source. NE-RCW36 PV diagram shows two separate velocity components, 4 and 7 \kms, which are also spatially separated. Since the column density calculation in this work is per spatial pixel, any spatially separable velocity structure does not affect the derivation of column density.

% CI PV maps + spectra

\ce{^{13}CO} generally appears to be more extensive than \ci\ as shown in Figure \ref{fig:mom0_maps}. However, we have determined that this is due the greater prevalence of artifacts in \ci\ maps, compared with those of \ce{^{13}CO} maps. The noise in both \ce{^{13}CO} and \ci\ maps is non-Gaussian, and is influenced in particular by fluctuations in weather at the times maps were taken. For \ci\ at the higher frequency of 492 GHz, maps are more affected by poor weather than \ce{^{13}CO} transition at 110 GHz. As the artifacts/noise are non-Gaussian, it has not been possible to determine a specific contour level of \ce{^{13}CO} below which \ci\ is no longer detected.

\subsection{Column densities}\label{sec:col}
The column density maps presented in this work is derived per spatial pixel, utilizing the integrated intensity maps in Section \ref{sec:maps_vel}. \ce{H2} column density is obtained from the optically thick \ce{^{12}CO}, \atci, and \ce{^{13}CO} column densities are corrected for opacity at each spatial pixel position.

For \ce{H2} column density, we utilize the empirical relation between \ce{H2} and \ce{^{12}CO} for Galactic molecular clouds \citep{Shetty2011},
\begin{equation}
  N_{\rm H_2}=2\times10^{20}\int T_{\rm MB}{\rm (CO,J=1\to0)}dV\,.
\end{equation}

We follow the column density calculation outlined in \citet{Oka2001} for \atci, which we repeat briefly here,
\begin{equation}
  N_{\rm CI}=1.98\times10^{15}\int T_{\rm MB}dV\,Q(T_{\rm ex})e^{E_1/(kT_{\rm ex})}\times\left[1-\frac{J_\nu(T_{\rm BB})}{J_\nu(T_{\rm ex})}\right]^{-1}\frac{\tau_{\rm CI}}{1-e^{-\tau_{\rm CI}}},
\end{equation}
where $\int T_{\rm MB}dV$ is the integrated emission of \ci, $Q(T_{\rm ex})$ is the partition function, 
\begin{equation}
    Q(T_{\rm ex})=1+3{\rm e}^{E_1/(kT_{\rm ex})}+5{\rm e}^{E_2/(kT_{\rm ex})}\,,
\end{equation}
with energy levels of $E_1/k=23.6$ K and $E_2/k=62.5$ K, $J_\nu(T_{\rm BB})$ and $J_\nu(T_{\rm ex})$ are the radiation temperature of cosmic background radiation ($T_{\rm BB}=2.7$ K) and excitation temperature ($T_{\rm ex}$) respectively, and $\tau_{\rm CI}$ is the opacity, 
\begin{equation}
  \tau_{\rm CI}=-{\rm ln}\left\{1-\frac{T_{\rm MB}}{\eta_f\left[J_\nu(T_{\rm ex})-J_\nu(T_{\rm BB})\right]}\right\}\,,
\end{equation}
here we assume the beam filling factor $\eta_f=1$. Assuming \atci\ has the same excitation temperature as the optically thick \ce{^{12}CO} at each of the spatial pixels, which is derived from the peak brightness temperature of \ce{^{12}CO} \citep{Glover2014}. We found the opacity of \atci\ is between 0.1 to 1.3 across the maps. The total column density of \atci\ is between $(0.1-8.4)\times10^{17}$  cm$^{-2}$, where IRAS16172--5028 has the highest peak column density of $8.4\times10^{17}$ cm$^{-2}$, and NE-RCW36 has the lowest peak column density of $3.3\times10^{17}$ cm$^{-2}$.

For \ce{^{13}CO} column density, we corrected \ce{^{13}CO} opacity by assuming that \ce{C^{18}O} is optically thin, and taking an isotopologue ratio of $\tau_{\rm 13CO}=7.4\tau_{\rm C18O}$ for the three sources in G333 GMC \citep{Wong2008}, and 5.5 for NE-RCW36. The opacity is then obtained by solving the brightness temperature-opacity relation,
\begin{equation}
  \frac{T_{\rm 13CO}}{T_{\rm C18O}}=\frac{1-e^{-\tau_{\rm 13CO}}}{1-e^{-\tau_{\rm C18O}}}.
\end{equation}
Similar to \atci, we assume \ce{^{13}CO} has the same excitation temperature as \ce{^{12}CO}, the total column density of \ce{^{13}CO} is then,
\begin{equation}
  N=\frac{8k\pi\nu^2}{hc^3g_{\rm u}A_{\rm ul}}\frac{\tau}{1-e^{-\tau}}e^{-E_u/(kT_{\rm ex})}Q(T_{\rm ex})\int T_{\rm b}dV\,,
\end{equation}
where $\int T_{\rm b}dV$ is the integrated emission of \ce{^{13}CO}, upper energy level of $E_u/k=5.3$ K, $A_{\rm ul}$ is the Einstein $A$ coefficient in s$^{-1}$, transition frequency $\nu$ in Hz, the degeneracy $g_u$, and $Q(T_{\rm ex})$ is the partition function \citep{Garden1991}. We found the opacity of \ce{^{13}CO} to be between 0.1 and 5.3, and column density in the range of $(0.1-100)\times10^{17}$ cm$^{-2}$. IRAS16172--5028 and IRAS16177--5018 have the highest peak column density ($\sim10^{19}$ cm$^{-2}$), while similar to \atci, NE-RCW36 has the lowest peak column density of $1.1\times10^{18}$ cm$^{-2}$.

Simulation and modeling on how well \atci\ traces molecular clouds suggest the $X_{\rm CI}-factor$ has a value of $1.1\times10^{21}$ cm$^{-2}$ K$^{-1}$ km$^{-1}$ s, an analogue to the widely used $X_{\rm CO}-factor$ that connects integrated \ce{^{12}CO} emission to \ce{H2} column density \citep[e.g.][]{Offner2014,Glover2014}. We apply this value to the integrated emission maps of \ci\ to derive \ce{H2} column density ($N_{\rm H2-CI}$) and compare this to the \ce{H2} column density maps we obtained from \ce{^{12}CO} maps ($N_{\rm H2-CO}$). We note that the conversion factor is obtained from simulation and may not be fully applicable to the observed region here, due to variables such as local abundance, chemical and physical conditions. We found $N_{\rm H2-CI}$ traces 80 to 100 per cent of $N_{\rm H2-CO}$ for regions where \ce{^{13}CO} column density is between $\sim5$ to $7\times10^{18}$ cm$^{-2}$ and \atci\ column density is between $\sim4$ to $8\times10^{17}$ cm$^{-2}$ for the three G333 {\it IRAS} sources, for NE-RCW36, the region is at $\sim1\times10^{17}$ cm$^{-2}$ for \ce{^{13}CO} and $\sim2$ to $3\times10^{17}$ cm$^{-2}$ for \atci. A summary of the derived physical properties are listed in Table \ref{tab:col}.

\begin{table}
%  \begin{center}
  \centering
  \caption{Observational and physical parameters of the sources.\label{tab:col}}
  \footnotesize
  \begin{tabular}{lcccc}
  \tableline
  \tableline
   & IRAS16164--5046$^a$ & IRAS16172--5028$^b$ & IRAS16177--5018$^c$ & NE-RCW36$^d$ \\
  \tableline
  Pointing center$^e$ ($\alpha,\delta$) & 16 22.17, -50 06.1 & 16 21.06, -50 35.38 & 16 21.54, -50 25.33 & 08 59.43, -43 43.87 \\
  Parent Complex & G333 & G333 & G333 & VMR-C \\
  Distance & 3.6 kpc$^f$ & 3.6 kpc$^f$ & 3.6 kpc$^f$ & 0.7 kpc$^g$\\
  rms chan$^{-1}$ & 0.4/0.3 K & 0.3/0.3 K & 0.3/0.3 K & 0.3/0.2 K\\
  (NANTEN2/Mopra) & &  &  &\\
  $V_{\rm CI}$ (\kms)$^h$ & $-47$, $-57$ & $-45$, $-51$, $-54$, $-61$ & $-45$, $-51$, $-58$ & 4.2, 7.0 \\
  $V_{\rm 13CO}$ (\kms)$^h$ & $-48$, $-57$ & $-45$, $-51$, $-54$, $-61$ & $-40$, $-43$, $-50$, $-56$ & 4.2, 7.0 \\
  $\Delta V_{\rm CI}$ (\kms)$^h$ & 2.5, 5.9 & 1.7, 5.6, 6.2, 4.4 & 6.7, 6.0, 6.5 & 2.5, 2.7 \\
  $\Delta V_{\rm 13CO}$ (\kms)$^h$ & 5.9, 5.6 & 2.8, 4.2, 6.2, 3.1 & 4.2, 3.1, 6.0, 6.8 & 1.5, 2.6 \\
  $\tau_{\rm CI}$$^j$ & 0.2 - 1.3 (0.5) & 0.2 - 0.8 (0.3) & 0.1 - 0.6 (0.3) & 0.1 - 0.8 (0.3) \\
  $\tau_{\rm 13CO}$$^j$ & 0.9 - 5.3 (2.8) & 0.1 - 5.3 (2.2) & 0.1 - 4.3 (2.2) & 0.04 - 3.0 (0.5) \\
  $N_{\rm CI}$ ($\times10^{17}$ cm$^{-2}$)$^j$ & 0.2 - 7 (2.2) & 0.9 - 8 (2.9) & 0.1 - 6 (2.6) & 0.1 - 3 (1.3) \\
  $N_{\rm 13CO}$ ($\times10^{17}$ cm$^{-2}$)$^j$ & 3 - 59 (9.6) & 0.3 - 99 (19) & 0.5 - 100 (28) & 0.1 - 11 (0.8) \\
  $T_{\rm ex}$ (K)$^j$ & 8 - 16 (12) & 14 - 29 (19)$^j$ & 17 - 29 (21) & 12 - 34 (19) \\
  $N_{\rm H2-CO}$ ($\times10^{22}$ cm$^{-2}$)$^k$ & 0.5 - 2.2 (1.9) & 2.4 - 6.6 (3.7) & 2.8 - 5.6 (3.9) & 0.7 - 2.1 (1.3) \\
  $N_{\rm H2-CI}$ ($\times10^{22}$ cm$^{-2}$)$^k$ & 0.3 - 2.6 (0.8) & 1.1 - 5.1 (1.8) & 0.8 - 4.1 (1.6) & 0.3 - 1.9 (0.9) \\
  \tableline
  \end{tabular}
  \tablenotetext{}{$^a$\textbf{IRAS16164--5046}: One of the most luminous far-infrared (FIR) sources in the Galaxy \citep{Becklin1973}; associated with the \hii\ region G333.6-0.22, H$_2$O and OH masers \citep{Batchelor1980,Caswell1998}; harbours a young OB cluster \citep[e.g.][]{Fujiyoshi2006,Grave2014}. $^b$\textbf{IRAS16172-5028}: Strongest source of SiO emission in G333 \citep{Lo2007}. Associated with \hii\ region G333.1-0.4, contains an embedded OB star cluster in very early evolutionary stages \citep{Figueredo2005}, and H$_2$O, OH and CH$_3$OH masers \citep{Breen2007,Caswell1998,Caswell1995}.    $^c$\textbf{IRAS16177--5018}: Part of the \hii\ region RCW106, contains the ultrahot star IRS1, likely an O3 I supergiant \citep{Roman_Lopes2009}.  $^d$\textbf{NE-RCW36}: Associated with the RCW36 \hii\ region, which contains a cluster of several hundred stars with the most massive star being a type O8 or O9 \citep[see][and references therein]{Minier2013}. $^e$The coordinates denote pointing centers for the maps (J2000). $^f$\citet{Lockman1979}. $^g$\citet{Murphy1991}. $^h$Velocity component ($V$) and line width ($\Delta V$) obtained from Gaussian fits to the spectra in Figure \ref{fig:pv_maps_spec}. $^j$Opacity ($\tau$), column density ($N$) of the \atci\ and \ce{^{13}CO} maps, excitation temperature ($T_{\rm ex}$) derived from peak \ce{^{12}CO} emission, $^k$\ce{H2} column density derived from \ce{^{12}CO} ($N_{\rm H2-CO}$) and \atci\ ($N_{\rm H2-CI}$). Note the ranges for $j$ and $k$ indicate the highest and lowest values across the maps of each of the sources (not per pixel), and median values are in brackets.}
%  \tablecomments{}
%  \end{center}
\end{table}

\section{Discussion}
Recent \atci\ mapping of the northern part of Orion-A GMC by \citet{Shimajiri2013} found a opacity of 0.1 - 0.75, similar to what we find (0.1 - 0.8) for three of our sources, with the exception of IRAS16164--5046 which reaches as high as 1.3 in optical depth at the \ci\ emission peak. The highest \atci\ column density in this work is of order of $\sim10^{17}$ cm$^{-2}$, which is an order of magnitude lower than the peak value $\sim10^{18}$ cm$^{-2}$ found in Orion-A. However, this could be due to effects such as resolution (i.e. beam filling factor less than 1). In fact, from modeling with the radiative transfer code RADEX \citep{VanderTak2007} for gas temperature shows $10^{18}$ cm$^{-2}$ fits our sources better (discussion later on).

\begin{figure}
  \epsscale{0.9}
  \plotone{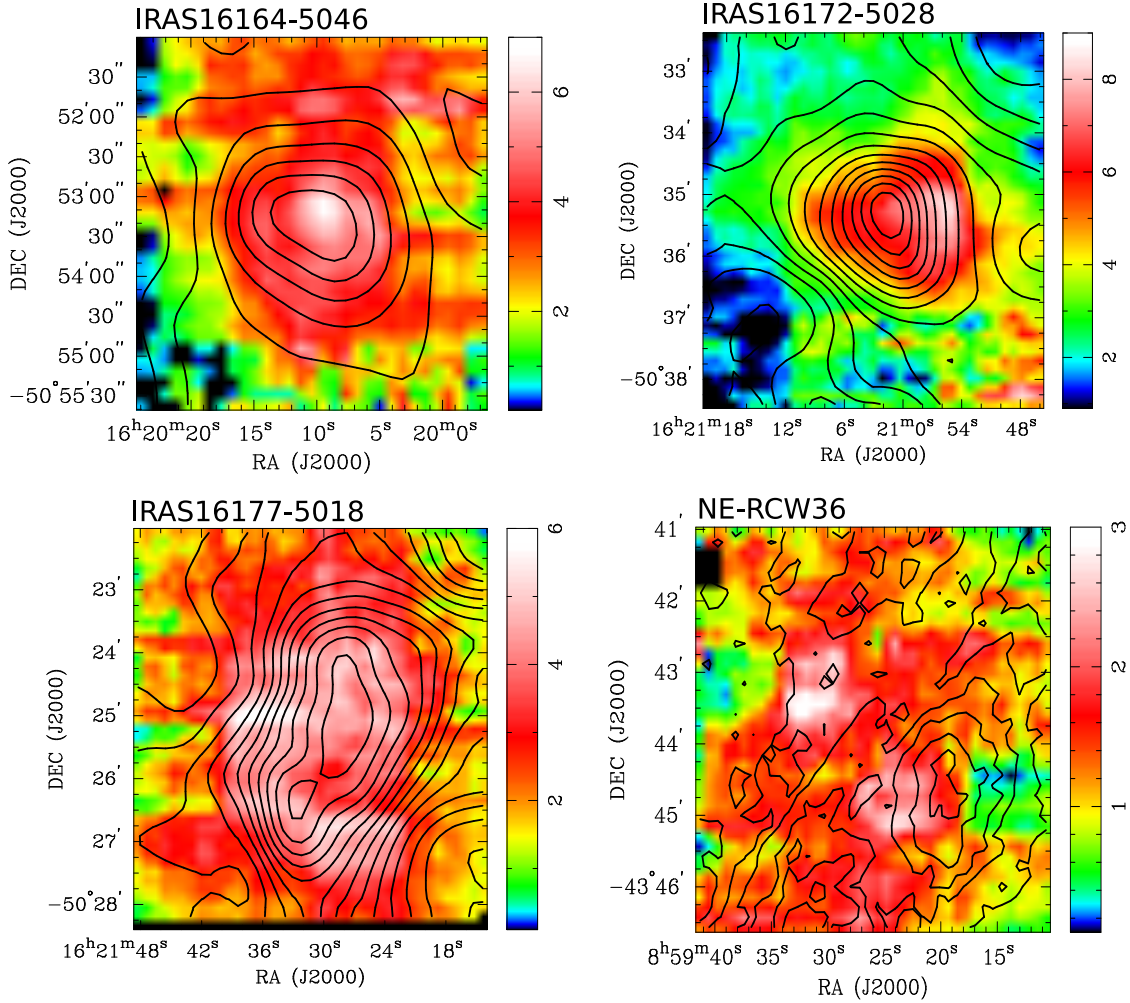}
%  \epsscale{0.48}
%  \plotone{f4a.eps} %iras16164_img_CI_H2_N.eps}
%  \plotone{f4b.eps} %iras16172_img_CI_H2_N.eps}
%  \plotone{f4c.eps} %iras16177_img_CI_H2_N.eps}
%  \plotone{f4d.eps} %rcw36_img_CI_H2_N.eps}
  \caption{\ce{H2} column density map (contours) overlaid on \atci\ column density map (color) of the four sources. The column density maps are at scale of $10^{21}$ cm$^{-2}$ and $10^{17}$ cm$^{-2}$ for \ce{H2} and \atci. The \ce{H2} contour levels start at 40 per cent of peak column density (Table \ref{tab:col}), except for IRAS16177--5018 which start at 50 per cent, the lowest visible contour level in the presented map. Contour steps are at 20$\sigma$, where 1$\sigma$ is $1\times10^{20}$ cm$^{-2}$ for IRAS16164--5046, IRAS16177--5018 and NE-RCW36, and $1.5\times10^{20}$ for IRAS16172-5028.\label{fig:N_maps}}
\end{figure}

We also find that the \atci\ column density peak is offset from the peak \ce{H2} column density when the excitation temperature (both quantities derived from the peak \ce{^{12}CO} emission, Section \ref{sec:col}) at the \ce{H2} peak column density  exceeds 25 K (Figure \ref{fig:N_maps}). This is unlikely to be due to the optical thickness of \ce{^{12}CO} `shifting' the apparent peak position, as from a cross check with the dust cores from ATLASGAL \citep{Csengeri2014} and BLAST \citep{Netterfield2009} data, their positions coincide well with the \ce{^{12}CO} peaks and thus the \ce{H2} column density peaks, except for source IRAS16177--5018, in which \ce{^{12}CO} is blended over various dust cores. In fact, at the peak \atci\ column density position, $T_{\rm ex}$ is within 15 to 20 K. The only source (IRAS16164--5046) that has both the \atci\ and \ce{H2} column density peaks coincide has an excitation temperature of $\sim15$ K at this position. Furthermore, if we double the excitation temperature ($>30$ K) in \atci\ column density calculations, it yields a lower column density at the \atci\ peak position. Modeling of \atci\ by \citet{Glover2014} shows around 80 per cent of \atci\ is found in regions with temperature below 30 K, and from our observations, we also find \atci\ is concentrated at places with lower excitation temperature (15 to 20 K). Depending on optical thickness, excitation temperature does not necessary equal gas kinetic temperature, and in the low opacity case, molecules are generally sub-thermally excited which makes the excitation temperature lower than kinetic temperature, so in either case, \atci\ is found mainly in low temperature gas. One possible explanation for this is that the \atci\ data presented here is from the observations of lower fine structure transition at 492 GHz with an energy level of 23 K. As the gas temperature rises, more of the neutral carbon atoms populate the higher transition level at 810 GHz with energy level of 62 K. To check whether this explanation is possible, we utilize the radiative transfer code RADEX by inputting gas temperatures in the range of 15 to 40 K. We find that as temperature rises, the intensity (population) of the 810 GHz transition increases from 2.8 to 20 K. Furthermore, the \ce{H2} column density we obtained with \atci\ is comparable to \ce{H2} column density derived from \ce{^{12}CO}, in which \ce{^{12}CO} is an unreliable tracer at high density as it becomes optically thick \citep[e.g.][]{Shetty2011}, and thus under estimates the column density. Despite that, the comparable \ce{H2} column density is suggesting that a portion of the neutral carbon atoms in the $^3P_1-^3P_0$ state appear to be missing. If this is the case, and if a substantial amount of atomic carbon is in an energy state above the ground state, we should see a position offset of the 810 GHz \ci\ transition for the sources here, and follow up observations in the future will further investigate this. More importantly, observations of both \ci\ transitions maybe necessary to recover the total \atci\ gas.

One possible uncertainty in our column density calculation comes from the velocity range in the integrated intensity maps. As shown in Section \ref{sec:maps_vel} these sources are not quiescent gases, they are turbulent with outflows (line wings). In order to access the effect of integrating over different gas components on column density derivation, two sets of velocity range are used: (1) integration over a velocity range that includes line wings/shoulders, (2) integration over the line width of the main core velocity component. In method (1), the \ce{H2} column density derived from \ci\ is comparable to those derived from \ce{^{12}CO} (Table \ref{tab:col}). With method (2), \ce{H2} column density obtained from \ci\ is approximately 10 to 30 per cent higher than those from \ce{^{12}CO}, but still within an order of magnitude. However, by integrating over the line width only can be problematic with \ce{^{12}CO}, as it is self-absorbed at the core velocity component, and the application of $X-factor$ requires \ce{^{12}CO} being optically thick and an integration over the emission range. Given the \ce{H2} column density derived from \ce{^{12}CO} and \ci\ in both methods are comparable and within the same order of magnitude, the choice of velocity range does not alter the conclusion in this work.

Another uncertainty in deriving \ce{H2} column density is the value of $X_{\rm CI}-factor$, unlike \ce{CO}, the relationship between \ci\ intensity and \ce{H2} column density is not well studied. The value we used in this work is from simulation studies \citep{Offner2014,Glover2014}, not specifically calibrated for the presented regions. The simulations do not take into account active star formation, where strong UV radiation increases the abundance of \atci, and thus, altering the value of $X_{\rm CI}-factor$. There is no observational/simulation study on $X_{\rm CI}-factor$ among active star forming regions (at the time of this work), but if star formation raises the abundance of \atci, then the $X_{\rm CI}-factor$ would be higher in these cases, and thus increases the derived \ce{H2} column density. We are in the process of completing the \atci\ mapping, along with various molecular lines at 3 mm wavelengths, we can calibrate the $X-factor$ of \atci\ and other molecules (e.g. \ce{HCO+}, \ce{CS}, \ce{HCN}) by comparing the \ce{H2} density estimated from SED fits of continuum emissions.

\section{Conclusion}
We present the first results of \atci\ mapping of VMR-C and G333 GMCs, comparing the column density of \ce{H2} derived from both \atci\ and \ce{CO} isotopologues. The \ci\ emission profile is similar to \ce{^{13}CO} with comparable line widths. We found \atci\ has opacity between 0.1 to 1.3, column density of $(0.1-8)\times10^{17}$ cm$^{-2}$, an order of magnitude lower than \ce{^{13}CO} column density. Utilizing the $X_{\rm CI}-factor$ we found an \ce{H2} column density of order $\sim10^{22}$ cm$^{-2}$, which is within the same order of magnitude of \ce{H2} column density derived from \ce{^{12}CO}, and near 100 per cent at \ci\ peak emission location. \ci\ emission tends to peak at regions with low gas temperature ($<20$ K), and it yields the same \ce{H2} column density as those derived from \ce{^{12}CO} at this temperature range. This could be due to part of the carbon atoms are in a higher excitation state, further mapping of \atci\ transition at 810 GHz will help confirm this.

Our results suggest if the gas is warm (above 25 K), it is recommended to observe both of the \ci\ transitions for a more accurate determination of \ce{H2} column density. \atci\ has the advantage of low opacity compared to \ce{^{12}CO}, while in low density and low extinction regions \ce{CO} is dissociated into neutral carbon and oxygen, making \atci\ a better tracer in these cases. However, in regions of dense gas where local abundance of \ce{^{13}CO} is known, \ce{^{13}CO} may be a better choice for probing \ce{H2} column density, as it is easily observed due to the lower frequency of its emission lines.


%% Included in this acknowledgments section are examples of the
%% AASTeX hypertext markup commands. Use \url without the optional [HREF]
%% argument when you want to print the url directly in the text. Otherwise,
%% use either \url or \anchor, with the HREF as the first argument and the
%% text to be printed in the second.

\acknowledgments

NL's postdoctoral fellowship is supported by CONICYT/FONDECYT postdoctorado, under project no. 3130540. LB acknowledges support by CONICYT Project PFB06. NANTEN2 Observatory is a collaboration between Nagoya University, Osaka University, KOSMA, Universit\"at zu K\"oln, Argelander-Institet Universit\"at Bonn, Seoul National University, ETH Z\"urich, University of New South Wales and Universidad de Chile. Mopra Telescope is part of the Australia Telescope and was funded by the Commonwealth of Australia for operation as National Facility managed by CSIRO until 2012.


%% in parentheses, after the keyword, but they will not be verified.

{\it Facilities:} \facility{Mopra}, \facility{NANTEN2}.

\end{document}